\documentclass[journal]{IEEEtran}
\usepackage{tikz}
\usepackage[sumlimits,intlimits]{amsmath}
\usepackage[english]{babel}
\usepackage{array}
\usepackage{tabularx}
\usepackage{amsfonts,amssymb}%
\usepackage{bm}%
\usepackage{graphicx,graphics}%
\usepackage[noadjust]{cite}%
\usepackage{color}%
\usepackage{url}
\usepackage{verbatim}
\usepackage{balance}
\usepackage{multirow}
\usepackage{multicol}
\usepackage{stfloats}
\usepackage{xspace}
\usepackage{enumerate}
\usepackage{rotating}
\usepackage[tight]{subfigure}
\usepackage{algorithm}
\usepackage{algorithmicx, algpseudocode}

\ifCLASSOPTIONpeerreview
\renewcommand{\baselinestretch}{2}
\fi

\begin{document}

\title {Cooperative  In-Vivo Nano-Network Communication at Terahertz Frequencies}

\author{
Qammer H. Abbasi,~\IEEEmembership{Senior Member,~IEEE,}
   Ali Arshad Nasir,~\IEEEmembership{Member,~IEEE,} Ke Yang, ~\IEEEmembership{Member,~IEEE,} Khalid
Qaraqe,~\IEEEmembership{Senior Member,~IEEE,} Akram
Alomainy,~\IEEEmembership{Senior Member,~IEEE}

\thanks{This publication was made possible by NPRP grant \# 7-125-2-061 from
the Qatar National Research Fund (a member of Qatar Foundation). The
statements made herein are solely the responsibility of the
authors.}
\thanks{Qammer and Khalid  are with the Dep. of Electrical and Computer Engineering, Texas A \& M University at Qatar. Qammer is also with Queen Mary University of London and UET, Lahore.; e-mail:
\{qammer.abbasi;k.qaraqe\}@tamu.edu.}
\thanks{Ali Arshad Nasir is King Fahd University of Petroleum and Minerals; e-mail:anasir@kfupm.edu.sa}
\thanks{ Ke Yang and A. Alomainy are with Antennas \& Electromagnetics School of Electronic Engineering and Computer Science
Queen Mary University of London, London; e-mail:
\{ke.yang;a.alomainy\}@qmul.ac.uk.}
 }

\maketitle
\vspace{-1cm}
\begin{abstract}
Nano devices have great potential to play a vital
role in future  medical diagnostics and treatment technologies
 because of its non-invasive nature and ability to reach
delicate body sites easily as compared to conventional
devices. In this paper, a novel concept of cooperative communication for in-vivo nano network is presented to enhance the communication among these devices. The effect on the system outage probability performance
is conducted for various parameters including relay placement, number of relays, transmit power, bandwidth and carrier frequency. Results show approximately a 10-fold increase in the system outage performance whenever an additional relay is included in the cooperative network, hence show a great potential of using cooperative communication to enhance the performance of nano-network at terahertz frequencies.
\end{abstract}
\begin{IEEEkeywords}
nano communication, Terahertz, body area network, channel modeling, cooperative communication.
\end{IEEEkeywords}

\section{Introduction}

Nano-technology has a critical role nowadays in  multidisciplinary domains including   biomedical, industrial control,  military and environment. However its impact in biomedical domain due to its non invasive nature
is making huge impact and driving research in this direction more intensively. Nano-network idea emerged from the connectivity of nano devices,  which was followed by  nano-communication proposal to enhance the features of these devices \cite{akyildiz2008nanonetworks}. Communication between nano-devices can be performed by different mechanism \emph{e.g.}, molecular, electromagnetic (EM), nanomechanical or acoustic \cite{andrew2000nanomedicine} \emph{etc}. However, EM based communication is considered as a feasible technique in the terahertz band  for exchange of  data  among these nano-machines \cite{Balasubramaniam2013} because biological tissues are non-ionized at these frequencies and also THz band is less susceptibility to scattering phenomenon \cite{piro2015terahertz}.

During past few years studies on body-centric communication have been gradually increasing
\cite{abbasibook,abbasi2012numerical}, however reduction in size requirement is making nano-technologies an
attractive choice for body-centric communication. Due to evolution of novel materials like graphene, \cite{luryi2013future} capable of operating at THz
frequencies, the interest in these frequencies for communication of nano-devices either on- or in-side the human body is growing as well. In addition, due to molecular resonance at these frequencies, even fine variation in water content or tissues can be detected, hence making this frequency propagation mechanism an emerging area of research for biomedical applications \cite{binzoni}. There are  numerous studies presented in literature discussing about the  applicability of THz
communication in  biomedical domain \cite{abbasireview,  berry2003optical, fitzgerald2003catalogue, jornet2011channel,akyildiz2010electromagnetic}. A detailed review on  the current state-of-the-art technologies and applicability of nano communication in biomedical application is presented by Abbasi \emph{et al.} in \cite{abbasireview}, while discussing about physical and networking layer concepts. The applicability of THz in nano-network is studied in \cite{akyildiz2010electromagnetic, jornet2011channel} and MAC-layer protocol is presented in \cite{jornet2012phlame}. Berry \emph{et al.} \cite{berry2003optical} characterised the optical properties of human tissues upto 2.5 THz. In order to further enhance the human tissue parameters at THz band, terhertz time domain spectroscopy is used to characterise the skin sample obtained from Blizard Institute in \cite{nishthz}. In this paper, collagen is created with different fibroblast cell numbers to see its effect on biological tissue parameters and signal propagation through the tissue. Ke \emph{et al.} \cite{ke} developed a tissue model and performed numerical and analytical studies inside the body to study pathloss and its relation with the dielectric loss.  In \cite{Qamthz} (authors of this paper), a novel channel model inside the body at THz frequencies for communication between these nano-devices based on extensive numerical and measurement studies have been presented, while considering various parameters like distance, number of sweat ducts and frequency. All of the above studies showed that, the communication range is very small at these frequencies because of high pathloss and in order to transmit reliable signal to receiver for getting useful information, a new paradigm of communication is needed. Although, as mentioned before there are some studies in open literature with regards to nano-communication, optical parameters characterisation, channel modeling and  applicability of THz communication in the biomedical domain, but there is no published research as per authors' knowledge about  the performance evaluation and advantage of using cooperative communication among these nano-devices.

\par In this paper, we are presenting for the first time in literature about the applicability and the study on the performance evaluation of cooperative in-vivo nano network communication at terahertz frequencies.  Outage probability performance evaluation is performed for various scenarios including different relay placement, \emph{i.e.,} relays placed vertically and randomly; different distances between the transmitter and the receiver, where relays were placed vertically and  variation in source power,  bandwidth and carrier frequency.
\begin{figure*}[t]
\centering
\includegraphics[width=0.8 \textwidth]{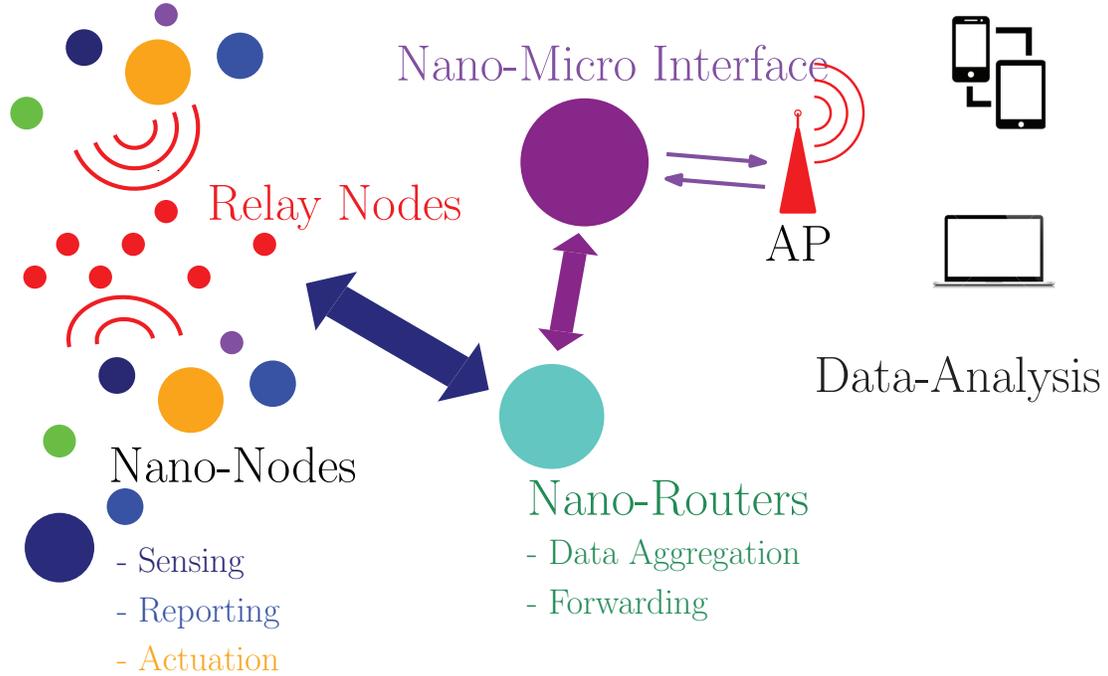}
\caption{Envisioned cooperative architecture for in-vivo nano-network } \label{figtt}
\end{figure*}

The rest of the paper is organized as follows. Section II presents the system model and simulation setup used in this study. Section III discusses the results for relay based communication considering various scenarios as mentioned above. Finally, conclusions are drawn in Section IV.

\section{In-Vivo Nano-nodes based System Model}
Body-centric nano-network can be categorised in three domains like traditional body-centric communication  \emph{i.e.,} on-body, off-body and in-body. An envisioned architecture \cite{akyildiz2010internet} for in-vivo nano-network  can be summarised as shown in Fig. \ref{figtt}.

 In Fig. \ref{figtt}, nano-nodes are the simplest and smallest nano-devices to perform simple computation and detection and then transmit it to relay nodes, which are also a small nano-devices and their task is  to do amplification and forwarding the received signal to  nano-routers.  These nano-devices (nano-routers) are slightly larger in terms of computational and behaviour capabilites  and can also act as a  control unit for the set of nano-nodes by ordering simple commands like read, sleep, wakeup \textit{etc}. They can be invasive or non-invasive depending on the application.  The nano-micro interface  is composed of hybrid devices which are used to exchange information between the two interfaces  and the last unit i.e., gateway allows the user to remotely control the system using internet.

\begin{figure}[!htpb]
\centering
\includegraphics[width=0.5\textwidth]{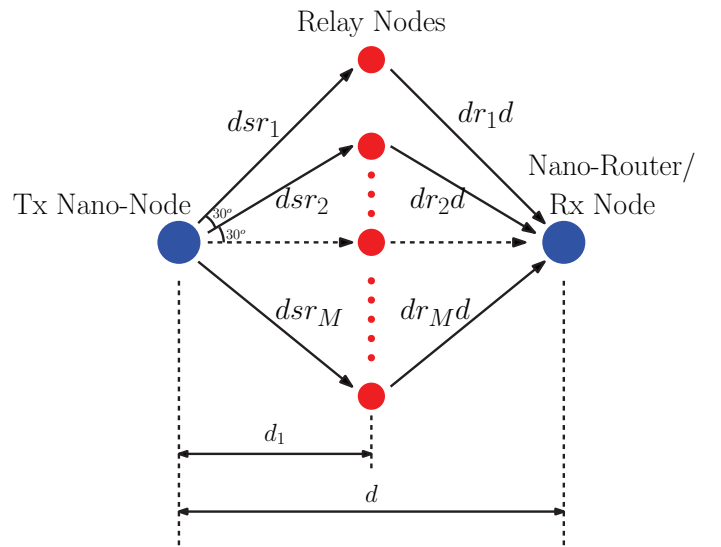}
\caption{System model for in-vivo cooperative communication at terahertz frequencies} \label{sys}
\end{figure}

As mentioned before in the introduction section, since pathloss at terahertz frequencies is considerably large even at very small distance and in order to transmit information to relatively larger distance in this paper  multiple relays are used to assist communication between the transmitter nano-nodes (Tx) and the nano-router/receiver node (Rx) as shown in Fig. \ref{sys} (\textit{Note that though Fig. 2 shows the vertical mid-way placement of relays, however, in Section III, we also analysed the performance of cooperative system with random relay placement}). To ensure implementation simplicity at the relays, an amplify-and-forward (AF) relaying assumption has been made.

Due to high pathloss, the direct link between the transmitter and the receiver would result in negligible received signal-to-noise ratio (SNR) at the receiver side and thus it can be ignored. In this paper  maximum ratio combining (MRC) has been employed at receiver side \emph{i.e.,} at nano-router \cite{WC}. Thus, the received SNR at the Rx,
\begin{figure*}[t]
    \centering
    \begin{minipage}[h]{0.48\textwidth}
    \centering
    \includegraphics[width=1.1\textwidth]{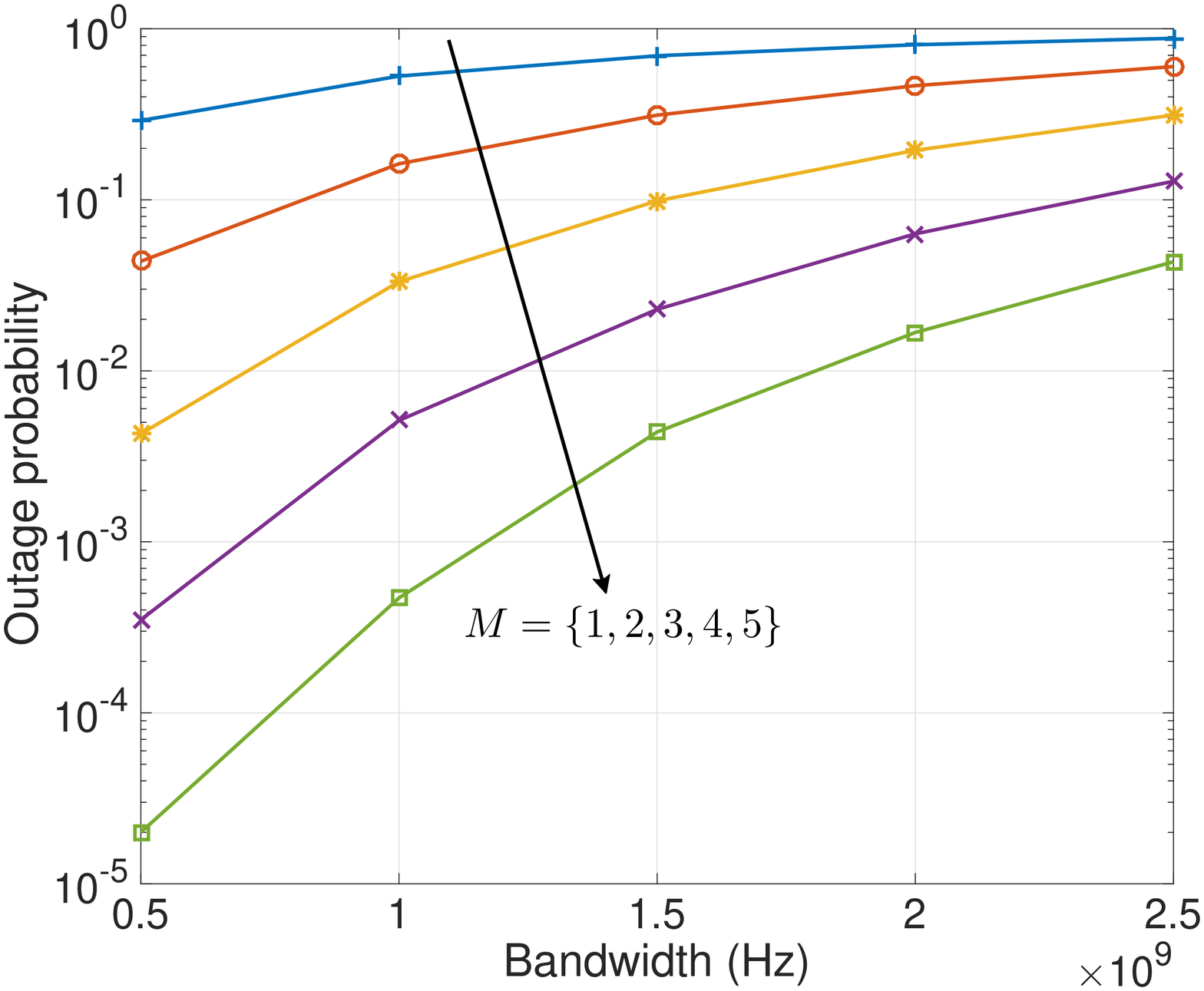}
  \caption{Outage probability performance for varying bandwidth, BW = $\{0.5,1,1.5,2,2.5\}$ GHz.}
  \label{BW}
  \end{minipage}
    \hspace{0.02cm}
    \begin{minipage}[h]{0.48\textwidth}
    \centering
    \includegraphics[width=1.1\textwidth]{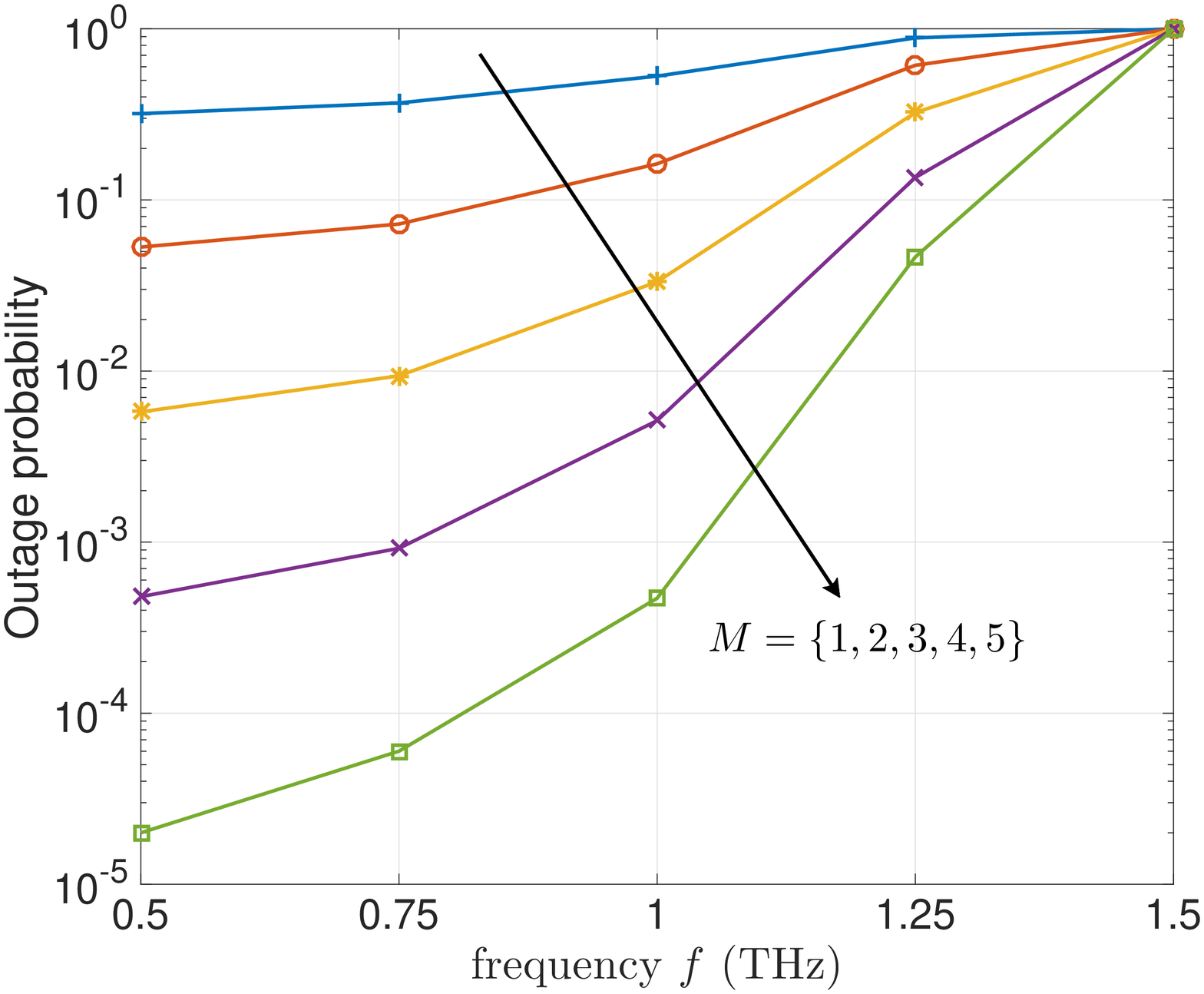}
  \caption{Outage probability performance for varying carrier frequency $f = \{0.5, 0.75, 1, 1.25,1.5\}$ THz.}
  \label{f}
  \end{minipage}
\end{figure*}
$\gamma_\text{Rx}$, through relaying link is given by \cite{Soliman-12-A}:
\begin{align}
\gamma_\text{Rx} = \sum_{i=1}^{M} \frac{\gamma_{s,r_i}
\gamma_{r_i,d} }{ \gamma_{s,r_i}  + \gamma_{r_i,d} + 1 },
\end{align}

where $\gamma_{s,r_i}  \triangleq \frac{ | h_{s,r_i} |^2 P_s }{ \ell_{s,r_i} \sigma^2 }$, is the received SNR at the $i_{th}$ relay due to the link between transmitter and the $i_{th}$ relay, $\gamma_{r_i,d}  \triangleq \frac{ | h_{r_i,d} |^2 P_\text{r} }{ \ell_{r_i,d} \sigma^2 }$ is the received SNR at the destination node due to the link between the $i_{th}$ relay and the Rx, $\ell_{s,r_i}$ and $\ell_{r_i,d}$ are the path losses from source transmitter to the $i_{th}$ relay and from the $i_{th}$ relay to the destination or Rx, respectively, $h_{s,r_i} \sim \mathcal{CN}(0,1)$ and $h_{r_i,d} \sim \mathcal{CN}(0,1)$ are complex normally distributed channel co-efficients for the respective links, respectively, $\sigma^2$ is the variance of additive white Gaussian noise, $P_s$ is the transmit power of source (transmitter) and $P_{r_i}$ is the transmit power of the $i$th relay. The path loss $\ell_{s,r_i}$ or $\ell_{r_i,d}$ at THz frequency by  Abbasi \emph{et al}.  as presented in  \cite{Qamthz} is used in this work and is given by:

\begin{eqnarray}\label{eq:pathloss}
 \ell_{s,r_i} &=& -0.2N + 3.98 + (0.44N + 98.48) d_{s,r_i}^{0.65}  \notag \\ && \hspace{2.3cm} + (0.068N+2.4)f^{4.07}  \\
\ell_{r_i,d} &=& -0.2N + 3.98 + (0.44N + 98.48) d_{r_i,d}^{0.65} \notag \\ &&  \hspace{2.3cm} +  (0.068N+2.4)f^{4.07},
\end{eqnarray}

where $N=5$ is the number of sweat ducts, f is frequency and $d_{s,r_i}$ and $d_{r_i,d}$ are the Euclidean distances from source transmitter to the $i_{th}$ relay and from the $i_{th}$ relay to the destination or Rx, respectively.

\underline{Performance Metric:}  \textit{Outage probability} is chosen as the performance metric to evaluate the communication performance of the considered system, where  outage probability in this paper is defined  as the probability that received SNR, $\gamma_\text{Rx} < \gamma_\text{th}$, where $\gamma_\text{th}$ is the threshold SNR for decoding the received signal.

\section{Discussion of Results}

In this work communication system for $M = \{1,\hdots,5$ relays is considered. The relays transmit with $100$ nW power, i.e., $P_{r_i} = - 40$ dBm (100 nW), $\forall$ $i \in M$. Noise variance per unit bandwidth (BW) is $ = -174$ dBm/Hz and threshold SNR is set to $\gamma_\text{th} = 10$ dB. Unless otherwise specified, the system bandwidth is set to be 1 GHz, carrier frequency to be $f = 1$ THz, the distance between the transmitter and the Rx to be $d = 0.2$ mm, and source transmit power is set to $P_s = - 40$ dBm (100 nW). Moreover, unless specified in particular,  all relays are placed vertically midway between the transmitter and the receiver, such that the two adjacent relays are $30^{\circ}$ angle apart from each other.

\begin{figure*}[!htpb]
    \centering
    \begin{minipage}[h]{0.48\textwidth}
    \centering
    \includegraphics[width=1.1 \textwidth]{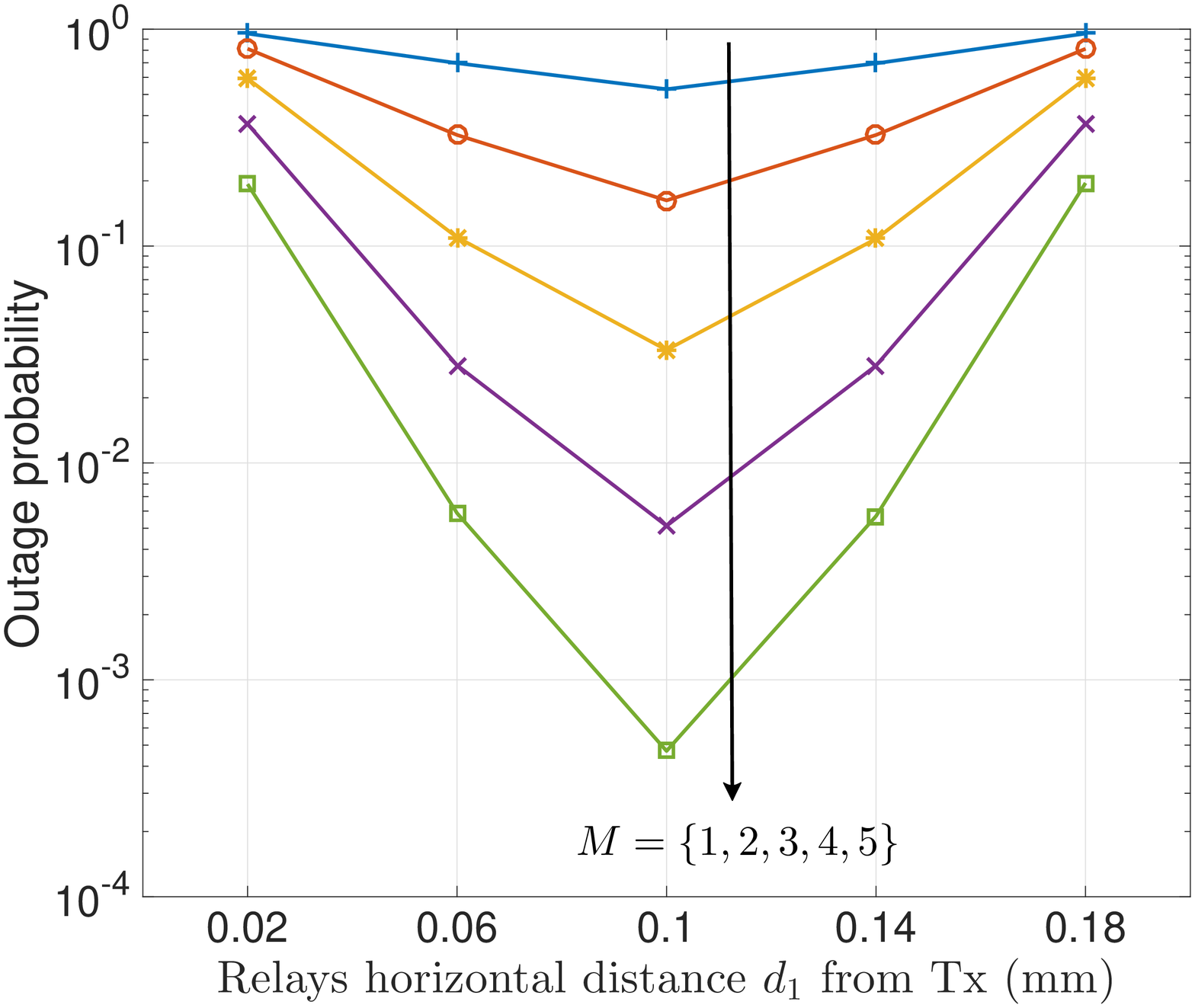}
  \caption{Outage probability performance for different placement of relays, i.e., relays placed vertically at a horizontal distance of $\{0.02,0.06,\hdots,0.18\}$ mm from the transmitter.}
  \label{dr}
  \end{minipage}
    \hspace{0.3cm}
    \begin{minipage}[h]{0.46\textwidth}
    \centering
    \includegraphics[width=1.1 \textwidth]{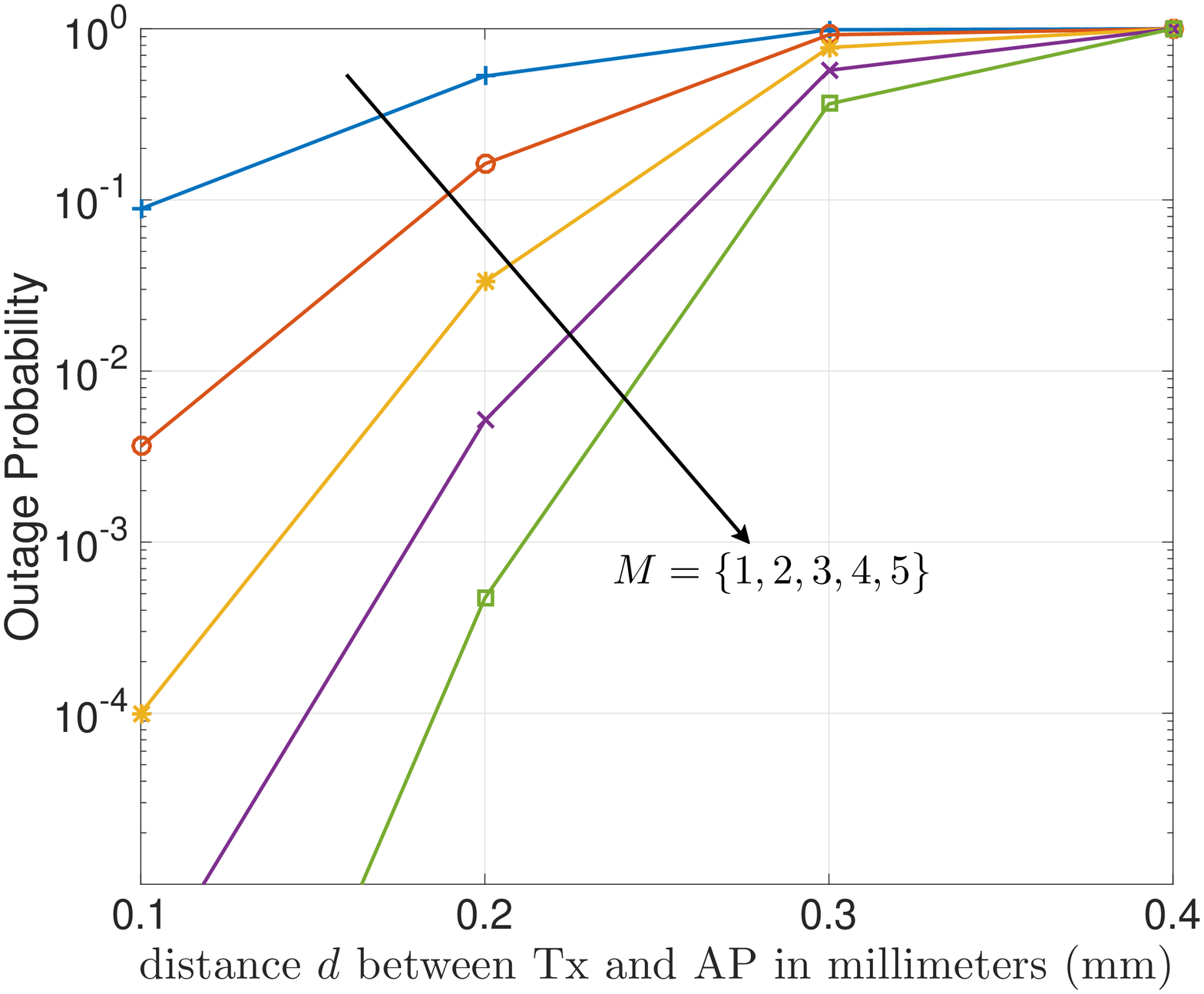}
  \caption{Outage probability performance for different distances between transmitter and the receiver, i.e., $d = \{0.1,0.2,0.3,0.4\}$ mm, where relays are placed vertically mid-way between transmitter and the receiver.}
  \label{dd}
  \end{minipage}
\end{figure*}

\begin{figure*}[!htpb]
    \centering
    \begin{minipage}[h]{0.48\textwidth}
    \centering
    \includegraphics[width=1.1 \textwidth]{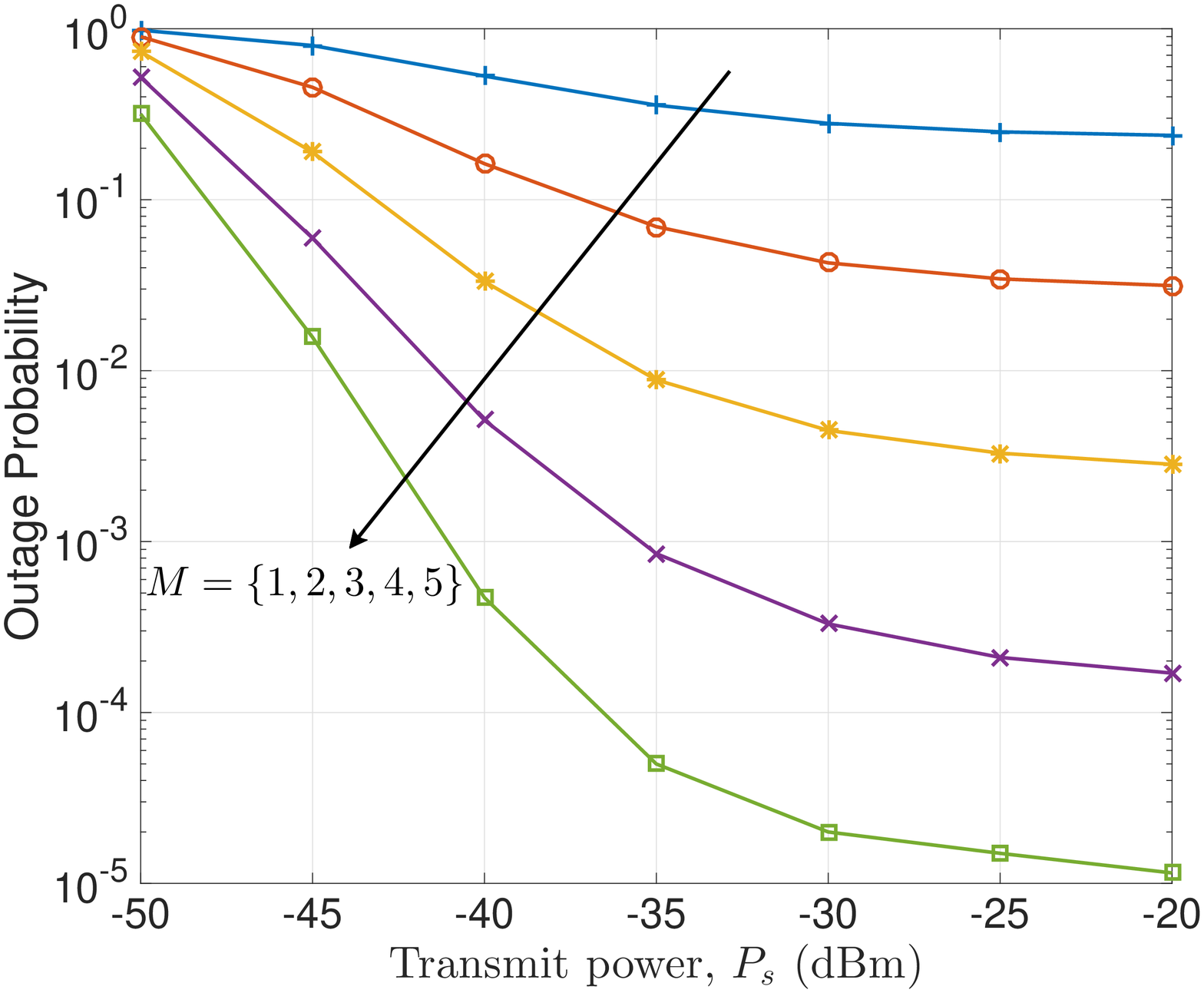}
  \caption{Outage probability performance for varying source transmit power, $P_s = \{-50,-45,\hdots,-20\}$ dBm.}
  \label{Ps}
  \end{minipage}
    \hspace{0.3cm}
    \begin{minipage}[h]{0.46\textwidth}
    \centering
    \includegraphics[width=1.1 \textwidth]{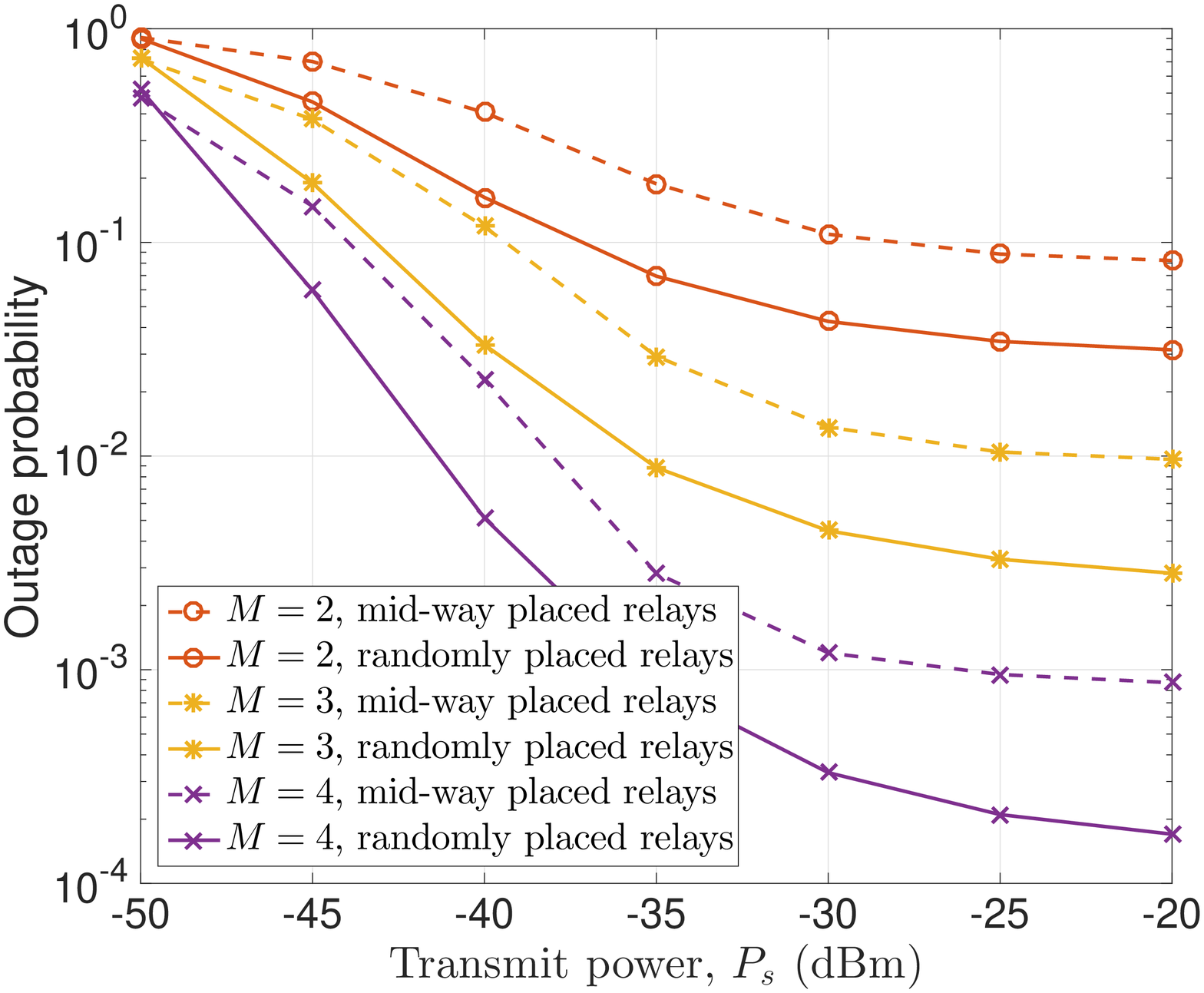}
  \caption{Outage probability performance for varying source power, $P_s = \{-50,-45,\hdots,-20\}$ dBm, while also considering random placement of relays within a circle of radius $\frac{d}{2}$ mm.}
  \label{Psr}
  \end{minipage}
\end{figure*}

Figs. \ref{BW} and \ref{f} demonstrate the outage probability performance for different values of bandwidth, BW = $\{0.5,1,1.5,2,2.5\}$ GHz and carrier frequency, $f = \{0.5, 0.75, 1, 1.25,1.5\}$ THz. It can be clearly observed that the outage probability performance is improved by increasing the number of relays. Fig. \ref{BW} shows that outage probability worsens by increasing the BW due to the increase in the noise variance, which decreases the received SNR $\gamma_\text{Rx}$. Similarly, an increase in the carrier frequency increases the path loss (as shown in  \eqref{eq:pathloss}), which decreases the received SNR $\gamma_\text{Rx}$. A poor outage probability performance for a single relay case is observed as shown in Figs. \ref{BW} and \ref{f}. However, it can be obviously improved by increasing the source and relay transmit powers from $100$ nW.

Fig. \ref{dr} shows the outage probability performance for different placement of relays, such that, relays are placed vertically at a horizontal distance of $\{0.02,0.06,\hdots,0.18\}$ mm from the transmitter. Fig. \ref{dr} shows that mid-way placement of the relays results in the minimum outage, whereas moving relays closer to the source or the Rx increases the outage probability. This is due to the increase in the path loss either for source-to-relays link or for relays-to-Rx link. Fig. \ref{dd} shows the outage probability performance for different distances between the transmitter and the receiver, i.e., $d = \{0.1,0.2,0.3,0.4\}$ mm, where relays are placed vertically mid-way between transmitter and the receiver. As expected, smaller the distance $d$, minimum is the outage due to minimum path loss. Fig. \ref{Ps} shows the outage probability performance for varying source transmit power, $P_s = \{-50,-45,\hdots,-20\}$ dBm $(10nW,\hdots,10 \mu W)$. The outage probability decreases with the increase in the source transmit power. Fig. \ref{Ps}  also shows that further increase in the source transmit power beyond $-30$ dBm ($1 \mu W$) results in marginal improvement in the outage probability. In Fig. \ref{Psr}, same results with some additional curves (dashed lines) are presented, which consider random placement of relays within a circle of radius $\frac{d}{2}$ mm. Fig. \ref{Psr} shows that random placement of relays worsens the outage probability performance, which is also supported by the findings of Fig. \ref{dr}.

\section{Conclusion}

In this paper, studies on  cooperative communication for  nano-scale electromagnetic based communication inside the body is presented. Various scenarios for relay based communication is considered including relay location, variation in source  power, bandwidth and carrier frequency variations \emph{etc}. Results highlight the increase in system performance by 10-fold when using cooperative communication hence  paving a new direction of research for applicability of single or multiple relay based communication in nano network for improving system performance.

\bibliographystyle{IEEEtran}
\bibliography{IEEEabrv,THz,review,chapter_3_ref,chapter_5_ref,chapter_4_ref,bibliography}

\end{document}